\documentclass[ecta,nameyear,draft]{econsocart}

\usepackage{amsmath,blkarray,bm}
\usepackage{amsfonts} 
\usepackage{amsthm} 
\usepackage{amssymb}
\usepackage{tikz}
\usepackage[mathscr]{euscript}
\usepackage{caption}
\usepackage{subcaption}
\usepackage{xcolor}
\usetikzlibrary{positioning}
\usepackage{pgfplots} \usepackage{pdfsync} \usepackage{xr} \usetikzlibrary{intersections}
\usepgfplotslibrary{fillbetween} 
\usepackage{centernot}
\usepackage{mathtools}
\usepackage{ stmaryrd }
\usetikzlibrary{shapes.geometric, arrows,positioning}
\usetikzlibrary{arrows}
\usepackage{bbm} \usepackage{natbib} \usepackage{setspace} \usepackage{enumitem}

\usepackage[flushleft]{threeparttable}

\usepackage{booktabs} 
\usepackage{tabularx} 

\definecolor{cornellred}{RGB}{179,27,27} 
\definecolor{cornellblue}{RGB}{55,135,176}
\definecolor{cornellgrey}{RGB}{96,94,92}

\RequirePackage[colorlinks,citecolor=blue,linkcolor=blue,urlcolor=blue,pagebackref]{hyperref}

\startlocaldefs

\numberwithin{equation}{section}
\theoremstyle{plain}
\newtheorem{theorem}{Theorem}[section] \newtheorem{proposition}{Proposition}[section] \newtheorem{lemma}{Lemma}[section]   \newtheorem{assumption}{Assumption}   
\newtheorem{definition}{Definition}[section]
\theoremstyle{definition}

\defcitealias{BMW}{BMW} 

\pgfplotsset{compat=1.18}

\endlocaldefs

\begin{document}

\begin{frontmatter}

\title{Testing Sign Congruence Between Two Parameters}

\begin{aug}
%
%
%
\author[id=au1,addressref={add1}]{\fnms{Douglas L.}~\snm{Miller}\ead[label=e1]{dlm336@cornell.edu}}
\author[id=au2,addressref={add2}]{\fnms{Francesca}~\snm{Molinari}\ead[label=e2]{fm72@cornell.edu}}
\author[id=au3,addressref={add2}]{\fnms{J\"{o}rg}~\snm{Stoye}\ead[label=e3]{stoye@cornell.edu}}
\address[id=add1]{%
\orgdiv{Department of Economics},
\orgname{Cornell University}}

\address[id=add2]{%
\orgdiv{Department of Economics},
\orgname{Cornell University}}
\end{aug}

\support{This version: July of 2025. We thank Hyewon Kim for excellent research assistance and the editor, two referees, as well as Gregory Cox, David Kaplan, Amanda Kowalski, and audiences at the 2023 Southern Economic Association Annual Meeting and the 2024 Econometric Society North American Winter Meeting for feedback.}
\begin{abstract}
We test the null hypothesis that two parameters $(\mu_1,\mu_2)$ have the same sign, assuming that (asymptotically) normal estimators $(\hat{\mu}_1,\hat{\mu}_2)$ are available. Examples of this problem include the analysis of heterogeneous treatment effects, causal interpretation of reduced-form estimands, meta-studies, and mediation analysis. A number of tests were recently proposed. We recommend a test that is simple and rejects more often than many of these recent proposals. Like all other tests in the literature, it is conservative if the truth is near $(0,0)$ and therefore also biased. To clarify whether these features are avoidable, we also provide a test that is unbiased and has exact size control on the boundary of the null hypothesis, but which  has counterintuitive properties and hence we do not recommend. We use the test to improve p-values in \citet{Kowalski} from information contained in that paper's main text and to establish statistical significance of some key estimates in \citet*{DippelEtAl2021}.\end{abstract}


\end{frontmatter}
\thispagestyle{empty}

\newpage

\section{Introduction}
It is common in empirical practice to investigate whether parameters of interest have the same sign.
Applications in the analysis of treatment effects loom large.
For example, researchers frequently ask whether an average treatment effect has the same sign across different subgroups of the population, as this knowledge helps in determining treatment protocols.
Researchers estimating marginal treatment effects or local average treatment effects may similarly investigate whether these effects are heterogeneous across groups.
Indeed, \citet*[BMW henceforth]{BMW} establish that testing whether the marginal treatment effect is constant across conditioning values of the propensity score can be expressed as a test of whether two parameters have the same sign,\footnote{These parameters are the difference in mean outcome for the treated group when a binary instrument takes value $1$ and value $0$ and the difference in mean outcome for the untreated group when a binary instrument takes value $1$ and value $0$.} and they implement a test of this hypothesis to analyze how family size affects children’s educational attainment.
\citet{Kowalski} carries out a similar test to determine whether ``women more likely to receive mammograms
experience higher levels of overdiagnosis.''
Even determining whether instrument validity fails in causal inference with non-compliers may include testing for sign congruence among population parameters \citep[Section 2.2]{Kim23}. 
Examples are not confined to the question of treatment effect heterogeneity.
The meta-analysis literature investigates whether two or more different studies estimate a treatment effect that has the significantly same sign \citep[see, e.g.,][]{ST22}. 
And in the rich literature on causal mediation \citep[see, e.g.,][for an overview]{mackinnon}, one asks whether the sign of the treatment effects into which the overall effect is decomposed are the same.
Instances of this testing problem are easy to find also outside the treatment evaluation literature: \citet{liu:mol24} show that a hypothesis of sign congruence is relevant in the analysis of the properties of an algorithmic fairness-accuracy frontier.

This paper addresses the problem of testing for sign congruence among two parameters as well as some extensions to related hypotheses.
Formally, we start with the null hypothesis
\begin{equation}\label{eq:simple_case}
H_0:\mu_1 \cdot \mu_2 \geq 0.
\end{equation}

We rediscover and recommend a deceptively simple test, which was proposed earlier by \citet{RCS93}.\footnote{We became aware of this literature reading \citet{Kim23} after independently deriving a solution to the problem. The test was also foreshadowed by the ``LR test'' in unpublished sections of \citet{KaplanDP}, a working paper precursor of \citet{KaplanJOE}.} In the special case \eqref{eq:simple_case} and if estimators $(\hat{\mu}_1,\hat{\mu}_2)$ are (asymptotically) normal with unit variances and uncorrelated (or positively correlated), the test rejects sign congruence at the $5\%$ significance level if $\hat{\mu}_1 \cdot \hat{\mu}_2 <0$ and furthermore $\min\{|\hat{\mu}_1|,|\hat{\mu}_2|\} >1.645$. 
The more general version of the test with unrestricted correlation is only marginally more complicated as it requires one to calibrate the critical value instead of using $1.645$. Nonetheless, the test not only is valid in the sense of controlling size uniformly over all $(\mu_1,\mu_2)$ on the null hypothesis, but is also uniformly more powerful than recently proposed tests that we discuss below. 
That said, it is still quite conservative near a true value of $(0,0)$ and has limited power against alternatives that are local to that point. We show that this feature can in principle be avoided by providing a test that has uniformly exact size control on the boundary of the null hypothesis and is also unbiased. However, that test rejects for some sample realizations that are arbitrarily close to the null hypothesis. Under an intuitive criterion that excludes such features and for a large range of values of an identified nuisance parameter, our recommended test has the largest possible rejection region (in the sense of set inclusion). 

A recent literature motivated by some of the above examples addressed the testing problem in \eqref{eq:simple_case} and put forward different testing procedures than the one we recommend here.
\citetalias{BMW} establish that \eqref{eq:simple_case} can be tested by simultaneously testing the sign of $\mu_1$ and $\mu_2$ and using Bonferroni correction to adjust for multiplicity of tests. They also consider a more general testing problem in which $(\mu_1,\mu_2)$ are hypothesized to lie in the union of two rotation symmetric origin cones whose vertices must be estimated. We generalize our test to that setting and formally show that the test we recommend is more powerful than the one in \citetalias{BMW}. \citet{ST22} use the same Bonferroni adjustment in the context of meta-analysis (though this application frequently calls for generalization to $K>2$ parameters, which we do not analyze).
\citet{Kowalski} proposes a heuristic bootstrap test and claims improved power over Bonferroni adjustment. We provide conditions under which these claims are true, verify that they hold in \citeauthor{Kowalski}'s (\citeyear{Kowalski}) setting, and clarify that this test is not valid without them.
Finally, the literature on causal mediation puts forward several testing approaches. Among these, \citet{montoya} perform the same bootstrap test as \citet{Kowalski}. 

\citet{Kim23} proposes a general testing procedure with uniform asymptotic validity that allows to test sign congruence among a finite number $K\ge 2$ of parameters when the asymptotic correlation matrix of the estimators of these parameters is unrestricted, along with a refinement to the testing procedure that may yield improved power performance for the case that $K>2$. In contrast, we only consider the case $K=2$ with arbitrary correlation.

The rest of this paper contains the following contributions. We restate the testing problem and briefly explain why it is difficult. We next provide our recommended test, which turns out to rediscover a test in \citet{RCS93}. We provide a simple independent proof of its validity and also establish the other, novel claims outlined above. We next clarify when the use of the heuristic bootstrap in \citet{Kowalski} and several other applications is justified. Specifically, such validity requires estimators to be (asymptotically) uncorrelated; but in this case, our recommended test is both easier to implement and uniformly more powerful. In general, the heuristic bootstrap test can have arbitrary size. We close with two empirical applications, both of which strengthen the statistical conclusions of published work. 

\section{Recommended Testing Procedure}

\subsection{Explaining the Testing Problem}

We start with a self-contained treatment of the problem of testing the null hypothesis in \eqref{eq:simple_case},
i.e., whether $\mu_1$ and $\mu_2$ have the same sign. For most of this paper, we impose:
\begin{assumption}\label{as:limit} We we have estimators $(\hat{\mu}_1,\hat{\mu}_2)$ distributed as
\begin{equation}\label{eq:normal_est}
\begin{pmatrix}\hat{\mu}_1 \\ \hat{\mu}_2 \end{pmatrix} \sim \mathcal{N}\left( \begin{pmatrix}\mu_1 \\ \mu_2 \end{pmatrix}, \begin{pmatrix}\sigma^2_1 & \rho\sigma_1\sigma_2 \\ \rho\sigma_1\sigma_2 & \sigma^2_2\end{pmatrix}\right),
\end{equation}
where $\sigma_1,\sigma_2>0$ and where $(\sigma_1,\sigma_2,\rho)$ are known. 
\end{assumption}
We think of this primarily as asymptotic approximation; see Section \ref{sec:finite} below for the analog of our main result when $(\sigma_1,\sigma_2,\rho)$ are consistently estimated using sample data.

This is a special case of a testing problem that has received attention in statistics; see \citet{Datta94} and references therein. It is difficult because, near a true parameter value of $(\mu_1,\mu_2)=(0,0)$, the natural ``plug-in'' test statistic $\hat{t} \equiv \frac{\hat{\mu}_1}{\sigma_1} \cdot \frac{\hat{\mu}_2}{\sigma_2}$ will be far from normally distributed. This is a red flag for applicability of common bootstrap schemes and the Delta method, and it is indeed known that both fail here.\footnote{With $\rho=\sigma_1=\sigma_2=1$, the example becomes Example 2.3 in \citet{Horowitz01}, which Horowitz uses to illustrate invalidity of the simple nonparametric bootstrap. In addition, a $\hat{t}$ computed from sample averages would converge at rate $n^{-1}$, raising another red flag for bootstrap applicability. 

The literature on statistics of this type goes back at least to \citet{Craig36}. In the context of causal mediation, Delta method inference was influentially advocated by \citet{Sobel1982}, although its invalidity is by now widely recognized in that literature.} 

\begin{figure}
\includegraphics[trim=4.5cm 8cm 4.5cm 8.8cm,clip=true,scale=.6]{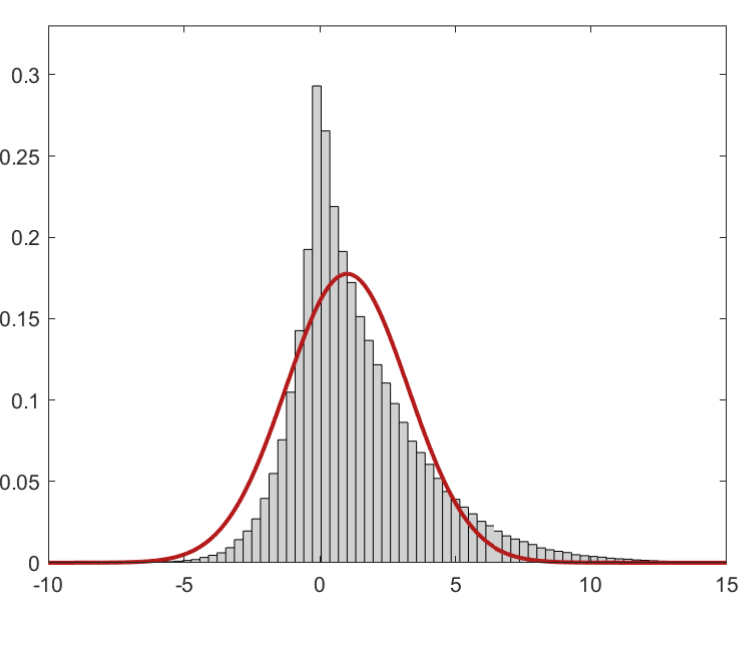}
\caption{True distribution (histogram) and delta method approximated distribution (curve) of $\hat{\mu}_1 \cdot \hat{\mu}_2$ if true parameter values are as in \eqref{eq:example}.}
\label{fig:delta}
\end{figure}

Because the Delta method works in a pointwise sense except literally at $(0,0)$, one might hope that the problem is ``exotic'' in the sense of being practically confined to a small region in parameter space. But this is not the case. Figure \ref{fig:delta} contrasts the true distribution of $\hat{\mu}_1 \cdot \hat{\mu}_2$ with an oracle Delta method approximation (i.e., using true parameter values) when
\begin{equation}\label{eq:example}
\begin{pmatrix}\hat{\mu}_1 \\ \hat{\mu}_2 \end{pmatrix} \sim \mathcal{N}\left( \begin{pmatrix}2 \\ 0.5 \end{pmatrix}, \begin{pmatrix}1 & ~~~.4 \\ .4 & ~~~1\end{pmatrix}\right).
\end{equation}
Here, one of the true means is two standard deviations away from zero, which one might intuit to be a ``safe'' distance, but the approximation is clearly  inaccurate.

An established literature in statistics considers general testing problems of this type and provides remedies that typically rely on pre-testing whether one is close to the point of irregularity and appropriately modifying critical values. 
Fortunately, much simpler tests are available for the special case of testing the null hypothesis in \eqref{eq:simple_case}. 
\citetalias{BMW} use a combination of Bonferroni and union-intersection arguments to show size control for the following test:
$$\text{Reject }H_0\text{ if }\hat{\mu}_1 \cdot \hat{\mu}_2<0 \text{ and }\min\{|\hat{\mu}_1|/\sigma_1,|\hat{\mu}_2|/\sigma_2\}\geq \Phi(1-\alpha/2).$$
In words, the estimated signs disagree and are individually significant at a level that reflects Bonferroni correction.

\citet{Kowalski} uses instead a heuristic test based on a simple nonparametric bootstrap. To translate this test to our setting, we equate the bootstrap and population distributions of estimation errors, thereby assuming perfect bootstrap approximation, possibly in some asymptotic limit.\footnote{The aforementioned use cases involve estimators for which validity of this approximation is essentially equivalent to asymptotic normality; for example, this follows for simple nonparametric or wild bootstrapping of sample moments from \citet{Mammen92}. We therefore do not think of bootstrap validity for the distribution of $(\hat{\mu}_1,\hat{\mu}_2)$ as a meaningful strengthening of our assumptions. Validity of simple percentile bootstrap and related approximations for $\hat{\mu}_1 \cdot \hat{\mu}_2$ does not follow nor, in fact, hold.} The test can then be written as follows.
\begin{align}
    \hspace{-.5cm}\text{Reject }H_0\text{ if }\Pr(\hat{\mu}_1^* \cdot \hat{\mu}_2^* >0)<\alpha,\text{ where }\begin{pmatrix}\hat{\mu}_1^* \\ \hat{\mu}_2^* \end{pmatrix} \sim \mathcal{N}\left( \begin{pmatrix}\hat{\mu}_1 \\ \hat{\mu}_2 \end{pmatrix}, \begin{pmatrix}\sigma^2_1 & \rho\sigma_1\sigma_2 \\ \rho\sigma_1\sigma_2 & \sigma^2_2\end{pmatrix}\right).\label{heur_boot}
\end{align}
 As with all bootstraps, the distribution of $(\hat{\mu}_1^*,\hat{\mu}_2^*)$ is conditional on the realized data; it is centered at the realized $(\hat{\mu}_1,\hat{\mu}_2)$.
The test's heuristic motivation is to check whether $0$ is inside a one-sided nonparametric percentile bootstrap confidence interval for $\mu_1 \cdot \mu_2$ that was computed in the spirit of \citet{Efron79}.

\subsection{Rediscovering a Simple Test}
It turns out that this problem has been previously considered in the statistics literature on \emph{qualitative interactions}, which proposes methods to test whether treatment effects' signs vary across groups of individuals. In particular, after developing a solution, we found that same solution in \citet[Theorem 6.1]{RCS93}, with related results for the case that the estimators are asymptotically uncorrelated in \citet[Theorem 1]{GS85}. While we provide a self-contained treatment that should be accessible to empirical economists, we do not claim novelty for results reported in this section.

With that said, we recommend the following test:
\begin{definition}[Recommended Test]\label{def:test}
\begin{align}
    \text{Reject }H_0 \text{ if }\hat{\mu}_1 \cdot \hat{\mu}_2<0 \text{ and }\min\{|\hat{\mu}_1|/\sigma_1,|\hat{\mu}_2|/\sigma_2\}\geq c_\alpha,\text{ where:}\label{eq:our_test}
\end{align}
\begin{itemize}
    \item[(i)] If $\rho \geq 0$, then $c_\alpha=\Phi(1-\alpha)$.
    \item[(ii)] Otherwise, $c_\alpha$ is characterized by
\begin{eqnarray}
    \sup_{\mu_2 \geq 0} \Pr \left(X_1 \cdot (\mu_2+X_2)<0,\min\{\left\vert X_1\right\vert,\left\vert\mu_2+X_2\right\vert\}> c_\alpha\right) = \alpha,\label{eq:def:c_alpha} \\
    \begin{pmatrix}X_1 \\  X_2 \end{pmatrix} \sim \mathcal{N}\left( \begin{pmatrix} 0 \\ 0 \end{pmatrix}, \begin{pmatrix}1 & ~~\rho \\ \rho & ~~1 \end{pmatrix}\right).\label{eq:def:c_alpha2}
\end{eqnarray}
\end{itemize}
\end{definition}
The critical value $c_\alpha$ is easily evaluated numerically. In fact, \citet[Table~\ref{table_results}]{GS85} provide a tabulation for several values of $\alpha$ (and several values of $K$, with $K$ the number of parameters whose sign congruence one is testing) assuming $\rho=0$; \citet[Table 2]{RCS93} provide a tabulation for several values of $\rho$ and $\alpha=0.05$ and $0.01$ for the case of two parameters. 

Notably, $\rho \geq 0$ includes the case where estimates derive from distinct subsamples and therefore $\rho=0$. The test then simplifies to
$$\text{Reject }H_0\text{ if }\hat{\mu}_1 \cdot \hat{\mu}_2<0 \text{ and }\min\{|\hat{\mu}_1|/\sigma_1,|\hat{\mu}_2|/\sigma_2\}\geq \Phi(1-\alpha).$$
For practical purposes, this simplification applies much more generally. Indeed, the statements $c_{.05}=1.645$ and $c_{.01}=2.326$ are correct up to all digits displayed whenever $\rho>-.8$; in other words, the critical values then coincide with one-sided ones up to typical recording accuracy (although not, at $\rho=-.8$, to numerical accuracy).\footnote{This relates to observations in \citet[][Table 2]{RCS93} but achieves higher accuracy. See Appendix \ref{app:critical_values} for more precise values and comparisons.}

In what follows, we call a test \emph{valid} if it controls size. Formally, let the sample statistic $\phi(\cdot)$ be an indicator of whether the test rejects. Then validity requires that 
\begin{align}
    \sup_{\mu_1,\mu_2:\mu_1 \mu_2 \geq0} E(\phi(\hat{\mu}_1,\hat{\mu}_2)) \leq \alpha.\label{eq:test_validity}
\end{align}
We call a test \emph{nonconservatively valid} if equality holds.
\begin{theorem}\label{th:main}
Suppose Assumption \ref{as:limit} holds. Then the test from Definition \ref{def:test} is nonconservatively valid.
\end{theorem}
\begin{figure}[t]
\tikzstyle{line} = [draw, -latex']
\begin{tikzpicture}[scale=0.9]
\linespread{0.9}
\path [fill=darkgray!30] (0,0) -- (0,4) -- (4,4)-- (4,0);
\path [fill=cornellred!15] (0,0) -- (4,0) -- (4,-1.65)-- (1.65,-1.65) -- (1.65,-4) -- (0,-4);
\draw[cornellred,thick] (4,-1.645)-- (1.645,-1.645) -- (1.645,-4);
\path [fill=cornellred!15] (0,0) -- (0,4) -- (-1.65,4)-- (-1.65,1.65) -- (-4,1.65) -- (-4,0);
\draw[cornellred,thick] (-1.645,4)-- (-1.645,1.645) -- (-4,1.645);
\path [fill=darkgray!30] (0,0) -- (0,-4) -- (-4,-4)-- (-4,0);
\draw[->] (-4,0) -- (4.1,0) node[below right] {$\hat \mu_1/\sigma_1$};
\draw[->] (0,-4) -- (0,4.1) node[above] {$\hat \mu_2/\sigma_2$};
\draw[->] (3,-.6) -- (3,-.01) ;
\draw[->] (3,-1.05) -- (3,-1.645) ;
\node at (3,-.82) {$1.645$};
\end{tikzpicture}
\caption{Visualization of null hypothesis (positive and negative orthant, gray) and recommended test's nonrejection region (null hypothesis plus additional red shaded area). For $\rho \geq 0$ and $\alpha=5\%$, the test's critical value is $\Phi^{-1}(.95) \approx 1.645$.}
\label{fig:test}
\end{figure}
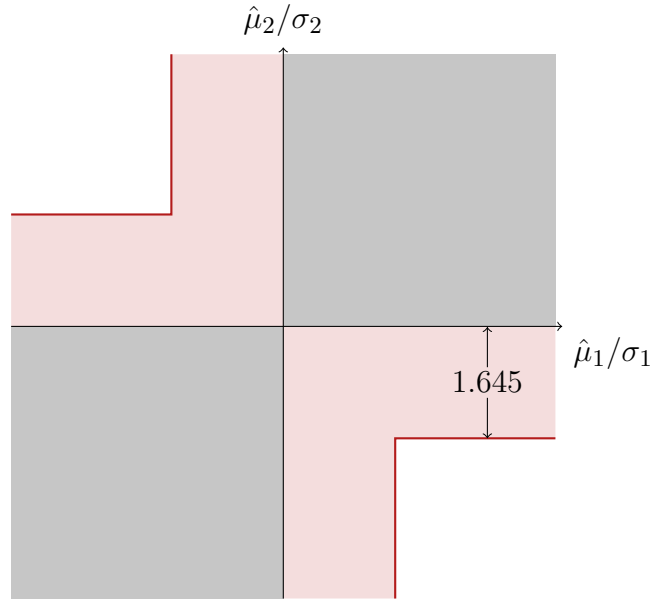
We provide an independent proof of this result in Appendix \ref{app:proof}. To build intuition, consider the visualization of the null hypothesis (gray) and nonrejection region (gray and red) of the test in Figure \ref{fig:test}. Suppose $\rho=0$ and $\sigma_1=\sigma_2=1$, so $(\hat{\mu}_1,\hat{\mu}_2)$ are bivariate standard normal but centered at $(\mu_1,\mu_2)$. Suppose also that $\alpha=.05$. The claim is that, for $(\mu_1,\mu_2)$ anywhere in the gray area, the probability of $(\hat{\mu}_1,\hat{\mu}_2)$ hitting the unshaded area is at most $\alpha$ and that this bound is attained in the worst case. To see that this is so, one can apply Anderson's Lemma separately on each ascending $45^\circ$ line to show that rejection probability is maximized on the boundary of $H_0$, i.e. for $(\mu_1,\mu_2)$ on some axis. Without further loss of generality, suppose $\mu_1=0$, i.e., the true parameter vector is on the vertical axis. Now imagine flipping the third and fourth quadrant in Figure \ref{fig:test}. If $\mu_1=0$, this does not affect the test's rejection probability. But the nonrejection region now contains all of $\{\hat{\mu}_1\geq -1.645\}$; hence, rejection implies rejection of the one-sided null that $\mu_1 \geq 0$ and so cannot exceed $5\%$.

The last argument can be extended to $\rho \geq 0$, but it is not available for $\rho<0$ and $c_\alpha$ must then be calibrated. It will approach $\Phi(1-\alpha/2)$ as $\rho \to -1$, explaining why a Bonferroni adjustment based test does not improve on $\Phi(1-\alpha/2)$ as critical value; in this sense, \citetalias{BMW}'s approach is tight. Finally, it is easy to verify that, for parameter values $(0,\mu_2)$ where $\mu_2 \to \infty$, the test is essentially a one-sided Wald test operated on $\hat{\mu}_1$. The test is therefore not conservative and approaches (although it may not attain) rejection probability of $\alpha$ on the null.

\subsection{Feasible Test}\label{sec:finite}
For simplicity and because it allows the concise statement of Proposition \ref{prop:monotonic} below, we mostly consider the idealized scenario of \eqref{eq:normal_est}. For completeness, here we clarify the feasible (assuming availability of a size $n$ sample), asymptotically valid version of the test.

A subtlety is that we want the test to be valid in a \emph{uniform} sense. As a reminder, pointwise validity would mean that, for fixed parameter values and as $n \to \infty$, a test controls size in the limit. This requirement is too weak here: It is fulfilled by the delta method except literally at $(\mu_1,\mu_2)=(0,0)$, but we saw that this finding is misleading for our purposes.\footnote{For more on why uniform inference may be necessary in certain settings, see \citet[][Section 3.1]{Canay_Shaikh_2017}.}  We, therefore, establish validity uniformly, including for $(\mu_1,\mu_2)$ ``close to'' $(0,0)$. To do so, we require, in Assumption \ref{as:asy} below, uniform convergence of the covariance matrix estimator and that a uniform central limit theorem holds; we refer to \citet{bug:can:shi15} for a discussion of low level conditions yielding these uniform convergence results.
\begin{assumption}\label{as:asy}
We have estimators $(\hat{\mu}_1,\hat{\mu}_2,\hat{\sigma}_1,\hat{\sigma}_2,\hat{\rho})$ s.t. the statements
\begin{gather}
\sqrt{n}\begin{pmatrix}\hat{\mu}_1-\mu_1 \\ \hat{\mu}_2-\mu_2 \end{pmatrix} \overset{d}{\to} \mathcal{N}\left( 0, \begin{pmatrix}\sigma^2_1 & \rho\sigma_1\sigma_2 \\ \rho\sigma_1\sigma_2 & \sigma^2_2\end{pmatrix}\right) \label{eq:asy_est1} \\
\frac{\hat{\sigma}_1}{\sigma_1} \overset{p}{\to}1,~~\frac{\hat{\sigma}_1}{\sigma_1} \overset{p}{\to}1,~~\hat{\rho}\overset{p}{\to}\rho \label{eq:asy_est2}
\end{gather}
hold uniformly across true values of $(\mu_1,\mu_2,\sigma_1,\sigma_2,\rho)$.
\end{assumption}
We then have the following definition and result.
\begin{definition}[Recommended Feasible Test]\label{def:feasible_test}
Reject $H_0$ if
 \begin{align}
    \hat{\mu}_1 \cdot \hat{\mu}_2<0 ~\text{  and  }~\sqrt{n}\min\{\lvert\hat{\mu}_1\rvert/\hat{\sigma}_1,\lvert\hat{\mu}_2\rvert/\hat{\sigma}_2\}\geq \hat{c}_\alpha,\label{eq:feasible_test}
\end{align}
where:
\begin{itemize}
    \item[(i)] If it is known that $\rho \geq 0$, then $\hat{c}_\alpha=\Phi(1-\alpha)$.
    \item[(ii)] Else, $\hat{c}_\alpha$ is characterized by
\begin{eqnarray}
    \sup_{\mu_2 \geq 0} \Pr \bigl(\hat{X}_1 \cdot (\mu_2+\hat{X}_2)<0,\min\{\lvert \hat{X}_1\rvert,\lvert\mu_2+\hat{X}_2\rvert\}> \hat{c}_\alpha\bigr) = \alpha,\label{def:feasible_c_alpha} \\
    \begin{pmatrix}\hat{X}_1 \\  \hat{X}_2 \end{pmatrix} \sim \mathcal{N}\left( \begin{pmatrix} 0 \\ 0 \end{pmatrix}, \begin{pmatrix}1 & ~~\hat{\rho} \\ \hat{\rho} & ~~1 \end{pmatrix}\right).\label{def:feasible_c_alpha2}
\end{eqnarray}
\end{itemize}
\end{definition}
\begin{proposition}\label{th:feasible}
Suppose Assumption \ref{as:asy} holds. Then the feasible test just defined is nonconservatively asymptotically valid, meaning that
\begin{align}
    \lim_{n \to \infty}\sup_{\mu_1,\mu_2:\mu_1 \mu_2 \geq 0} \Pr(\text{\eqref{eq:feasible_test} obtains}) =\alpha.\label{eq:feasible_test_valid}
\end{align}
\end{proposition}
The limit in \eqref{eq:feasible_test_valid} is taken outside the supremum, establishing uniform validity. Again, switching the order would lead to a much weaker and, in this context, insufficient claim.

For the remainder of this paper, we return to the idealized setting of Assumption \ref{as:limit}.

\subsection{Validity of the Heuristic Bootstrap}
Validity of the heuristic bootstrap for this testing problem is far from obvious. Indeed, recall that the test is based on a simple nonparametric bootstrap approximation, but such an approximation presumes symmetry of the distribution that is being simulated \citep[as nicely discussed in][Section 10.16]{Hansen}. Figure \ref{fig:delta} therefore casts doubt not only on the delta method but also on inference based on the percentile bootstrap. Indeed, we can report the following result.

\begin{proposition} \label{prop:heuristic}
    Without further restrictions on $(\rho,\mu_1,\mu_2)$, the true rejection probability of the heuristic bootstrap test can take any value in $(0,1)$, irrespective of whether $H_0$ is true or not.
\end{proposition}

This raises the question whether the heuristic bootstrap is valid in applications where it has previously been used. Fortunately, we are able to (i) establish such validity for applications where $\rho=0$ is known and (ii) verify that this case obtains in \citet{Kowalski}. The first of these observations follows from the next result.

\begin{proposition}\label{prop:compare}
For $\rho=0$, the \citetalias{BMW} rejection region is contained in the interior of the heuristic bootstrap rejection region, which in turn is in the interior of the recommended test's rejection region. 
\end{proposition}

Given validity of the recommended test, the proposition establishes a fortiori that, with $\rho=0$, the heuristic bootstrap is valid in the sense of controlling size. This case is not without interest as it obtains when estimates $(\hat{\mu}_1,\hat{\mu}_2)$ come from independent samples or distinct subsamples of the same (exchangeable) sample. For example, this characterizes the application to meta-analyses because independent studies presumably have independent errors. Much less obviously, it is also true in \citet{Kowalski}. Formally, by tedious manipulation of some expressions in  \citet{Kowalski}, we can show the following:

\begin{proposition}\label{prop:kowalski}
    The testing problem in \citet[Section 4.2]{Kowalski} is characterized by $\rho=0$.
\end{proposition}

Taken together, these propositions clarify that all claims \citet{Kowalski} makes about the test for her setting are correct; in particular, the test is both valid and less conservative than Bonferroni adjustment. However, this assessment depends on $\rho=0$. While we prove that this holds in her setting, it will not extend to more general estimators of treatment effects across groups and over time, or in the context of mediation analysis. Furthermore, the heuristic bootstrap's appeal is limited by the fact that, if it is valid, then a simpler and yet uniformly more powerful test is available.

\section{Can the Test be Further Improved?}
While our recommended test improves on some other tests in the literature, its power is low around $(0,0)$ -- for example, for known parameter values $(\sigma_1,\sigma_2,\rho)=(1,1,0)$, power at $(0,0)$ is easily computed to be $\alpha/10$, where $\alpha$ is nominal rejection probability. By the same token, the test is biased -- close to $(0,0)$ but outside of $H_0$, rejection probability is lower than at $(\mu,0)$ for large $\mu$, even though the latter is in $H_0$. Can we remedy these issues? It turns out that the answer is ``yes'' if we allow for highly uncoventional tests that we do not seriously recommend. We then impose a condition that excludes such tests and show that, under this condition, the answer is ``no'' and, for most values of $\rho$, the proposed test is in a strong sense best.

Let $\rho=0$ and $\alpha=.05$. Partition the positive real line into bins with end points
\begin{multline*}
  (\Phi^{-1}(.5),\Phi^{-1}(.55),\Phi^{-1}(.6),\Phi^{-1}(.65),\ldots,\Phi^{-1}(.9),\Phi^{-1}(.95)) \\ \approx(0,0.13,0.25,0.39,0.52,0.67,0.84,1.04,1.28,1.645), 
\end{multline*}
where $\Phi^{-1}(\cdot)$ is the standard normal quantile function. In words, the positive real line is divided into intervals that correspond to ``$.05$-quantile bins'' of the standard normal distribution. We can then define the following test, which is visualized in Figure \ref{fig:fractal}:
\begin{align}
    \text{Reject }H_0\text{ if }\hat{\mu}_1 \cdot \hat{\mu}_2<0 \text{ and }(|\hat{\mu}_1|/\sigma_1,|\hat{\mu}_2|/\sigma_2)\text{ are in the same bin.}\label{eq:fractal_test}
\end{align}
A similar test was considered in \citet{Berger89}  (and criticized in \citet{PearlmanWu}). In more closely related work, \citet{vangarderen2022} develop the same idea for $H_0:\mu_1 \cdot \mu_2 = 0$ and so do not encounter a question of unbiasedness, but they do show essential uniqueness of this test. Numerical computation of nonlinear rejection regions with similar intuitions was proposed (at least) in \citet{chiburis}, \citet{KaplanDP}, and \citet{HGG}. 

We next show that the test in Eq.~\eqref{eq:fractal_test} is asymptotically similar (its rejection probability equals $0.05$ everywhere on the boundary of $H_0$) and unbiased ($H_0$ is a lower contour set of the power function).

\begin{figure}[t]
\tikzstyle{line} = [draw, -latex']
\begin{tikzpicture}[node distance = 2cm, auto]
\path [fill=darkgray!30] (0,0) -- (0,4) -- (4,4)-- (4,0);
\path [fill=cornellred!15] (4,-1.645)-- (1.645,-1.645) -- (1.645,-1.28) -- (1.28,-1.28) -- (1.28,-1.04) -- (1.04,-1.04) -- (1.04,-0.84) -- (0.84,-0.84) -- (0.84,-0.67) -- (0.67,-0.67) -- (0.67,-0.52) -- (0.52,-0.52) -- (0.52,-0.39) -- (0.39,-0.39) -- (0.39,-0.25) -- (0.25,-0.25) -- (0.25,-0.13) -- (0.13,-0.13) -- (0.13,0) -- (4,0);
\draw[cornellred,thick] (4,-1.645)-- (1.645,-1.645) -- (1.645,-1.28) -- (1.28,-1.28) -- (1.28,-1.04) -- (1.04,-1.04) -- (1.04,-0.84) -- (0.84,-0.84) -- (0.84,-0.67) -- (0.67,-0.67) -- (0.67,-0.52) -- (0.52,-0.52) -- (0.52,-0.39) -- (0.39,-0.39) -- (0.39,-0.25) -- (0.25,-0.25) -- (0.25,-0.13) -- (0.13,-0.13) -- (0.13,0) -- (0,0);
\path [fill=cornellred!15] (1.645,-4)-- (1.645,-1.645) -- (1.28,-1.645) -- (1.28,-1.28) -- (1.04,-1.28) -- (1.04,-1.04) -- (0.84,-1.04) -- (0.84,-0.84) -- (0.67,-0.84) -- (0.67,-0.67) -- (0.52,-0.67) -- (0.52,-0.52) -- (0.39,-0.52) -- (0.39,-0.39) -- (0.25,-0.39) -- (0.25,-0.25) -- (0.13,-0.25) -- (0.13,-0.13) -- (0,-0.13) -- (0,-4);
\draw[cornellred,thick] (1.645,-4)-- (1.645,-1.645) -- (1.28,-1.645) -- (1.28,-1.28) -- (1.04,-1.28) -- (1.04,-1.04) -- (0.84,-1.04) -- (0.84,-0.84) -- (0.67,-0.84) -- (0.67,-0.67) -- (0.52,-0.67) -- (0.52,-0.52) -- (0.39,-0.52) -- (0.39,-0.39) -- (0.25,-0.39) -- (0.25,-0.25) -- (0.13,-0.25) -- (0.13,-0.13) -- (0,-0.13) -- (0,0);
\path [fill=cornellred!15] (-4,1.645)-- (-1.645,1.645) -- (-1.645,1.28) -- (-1.28,1.28) -- (-1.28,1.04) -- (-1.04,1.04) -- (-1.04,0.84) -- (-0.84,0.84) -- (-0.84,0.67) -- (-0.67,0.67) -- (-0.67,0.52) -- (-0.52,0.52) -- (-0.52,0.39) -- (-0.39,0.39) -- (-0.39,0.25) -- (-0.25,0.25) -- (-0.25,0.13) -- (-0.13,0.13) -- (-0.13,0) -- (-4,0);
\draw[cornellred,thick] (-4,1.645)-- (-1.645,1.645) -- (-1.645,1.28) -- (-1.28,1.28) -- (-1.28,1.04) -- (-1.04,1.04) -- (-1.04,0.84) -- (-0.84,0.84) -- (-0.84,0.67) -- (-0.67,0.67) -- (-0.67,0.52) -- (-0.52,0.52) -- (-0.52,0.39) -- (-0.39,0.39) -- (-0.39,0.25) -- (-0.25,0.25) -- (-0.25,0.13) -- (-0.13,0.13) -- (-0.13,0) -- (0,0);
\path [fill=cornellred!15] (-1.645,4)-- (-1.645,1.645) -- (-1.28,1.645) -- (-1.28,1.28) -- (-1.04,1.28) -- (-1.04,1.04) -- (-0.84,1.04) -- (-0.84,0.84) -- (-0.67,0.84) -- (-0.67,0.67) -- (-0.52,0.67) -- (-0.52,0.52) -- (-0.39,0.52) -- (-0.39,0.39) -- (-0.25,0.39) -- (-0.25,0.25) -- (-0.13,0.25) -- (-0.13,0.13) -- (0,0.13) -- (0,4);
\draw[cornellred,thick] (-1.645,4)-- (-1.645,1.645) -- (-1.28,1.645) -- (-1.28,1.28) -- (-1.04,1.28) -- (-1.04,1.04) -- (-0.84,1.04) -- (-0.84,0.84) -- (-0.67,0.84) -- (-0.67,0.67) -- (-0.52,0.67) -- (-0.52,0.52) -- (-0.39,0.52) -- (-0.39,0.39) -- (-0.25,0.39) -- (-0.25,0.25) -- (-0.13,0.25) -- (-0.13,0.13) -- (0,0.13) -- (0,0);
\path [fill=darkgray!30] (0,0) -- (0,-4) -- (-4,-4)-- (-4,0);
\draw[->] (-4,0) -- (4.1,0) node[below right] {$\hat \mu_1/\sigma_1$};
\draw[->] (0,-4) -- (0,4.1) node[above] {$\hat \mu_2/\sigma_2$};
\draw[->] (3,-.6) -- (3,-.01) ;
\draw[->] (3,-1.05) -- (3,-1.645) ;
\node at (3,-.82) {$1.645$};
\end{tikzpicture}
\caption{A test with uniformly exact size control (i.e., constant rejection probability of $5\%$) on the boundary of $H_0$. Here, $H_0$ is gray (positive and negative quadrant) and the nonrejection region beyond $H_0$ is red (irregularly shaped object in second and fourth quadrants).} \label{fig:fractal}
\end{figure}
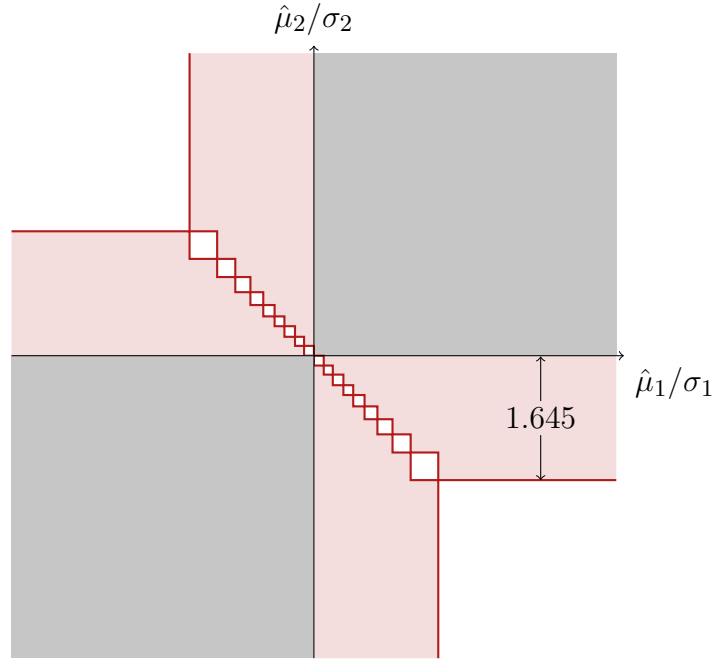
\begin{proposition}\label{prop:fractal}
    Let $\rho=0$. The test in Eq.~\eqref{eq:fractal_test} has rejection probability equal to $0.05$ at every point $(\mu_1,\mu_2)$ with $\mu_1 \cdot \mu_2=0$, i.e., uniformly on the boundary of $H_0$. The test is furthermore unbiased. Indeed, its rejection probability strictly exceeds $0.05$ everywhere outside $H_0$ and is strictly below $0.05$ in the interior of $H_0$.
\end{proposition}

Proposition \ref{prop:fractal} establishes that, in an abstract statistical sense, the test has attractive properties. However, even disregarding that its construction relies on $\alpha=.05$ evenly dividing into $1$, the test is not intuitively compelling by any stretch. For example, it rejects for $(\hat{\mu}_1,\hat{\mu}_2)=(m,-m)$ no matter how small $m$ is. This means that the data may be arbitrarily close to $H_0$ and still be deemed to reject it; by the same token, there are points in the rejection region where the Wald test p-value against the simple null hypothesis that $(\mu_1,\mu_2)=(0,0)$ is arbitrarily close to $1$. In addition, there are fixed alternatives against which the test is not consistent, i.e. it does not asymptotically reject them with certainty.\footnote{These observations are related to more general points made in \citet{Andrews2012} and \citet{PearlmanWu}.}

These considerations lead us to not recommend the test. However, they show that our proposed test can be most powerful only within a restricted class of tests even if basic desiderata like nontriviality (the test in fact depends on data) are imposed, and also that unbiasedness (the null hypothesis is a lower contour set of the power function) may be an unrealistic demand.
We next argue that the recommended test is best within a reasonable comparison set. To this end, consider the following definition.\begin{definition}\label{def:monotonic_test}
    A test is monotonic if, for any $(m_1,m_2)$ in the rejection region, one cannot leave the rejection region by increasing the absolute value of one or both of $(m_1,m_2)$ while leaving their signs unchanged.
\end{definition}
In another belatedly discovered precedent, \citet[][Section 4.2]{KaplanDP} proposes the same monotonicity condition and establishes that it brings one back to the recommended test for $\rho=0$, a finding that we will generalize below. Intuitively, if a test is monotonic, one cannot pass from the rejection region into the nonrejection region by moving \emph{away from} the null hypothesis. For example, in Figure \ref{fig:test}, any ``forbidden'' movement would go southeast [northwest] from within the rejection region's southeastern [northwestern] component. The condition is evidently met. In contrast, the square components of the rejection region in Figure \ref{fig:fractal} violate it. We find this condition appealing.\footnote{This appeal may relate to the fact that, along movements described in Definition \ref{def:monotonic_test}, the maximal likelihood of the data under the null hypothesis cannot increase. Note however that, while the rejection region in Figure \ref{fig:test} is a lower contour set of this maximized likelihood, Definition \ref{def:monotonic_test} allows for rejection regions that are not. Indeed, it is easy to see that any rejection region that is a lower contour set of the maximized likelihood corresponds to our recommended test except for possibly using a different critical value. See \citet{Kline11} for several other examples of counterintuitive nonmonotonicities in LR-type test statistics and, by implication, rejection regions.} That said, it excludes tests that were recommended in the aforementioned literature, and users who disagree with its heuristic appeal should certainly feel free to use such tests.

We next establish an important property of monotonic tests.\begin{proposition}\label{prop:monotonic}
    If a monotonic test is valid in the sense of \eqref{eq:test_validity}, then its rejection region is contained in $$\bm{R}_\alpha := \{\hat{\mu}_1,\hat{\mu}_2~:~\hat{\mu}_1 \cdot \hat{\mu}_2 <0,~\min\{\left\vert\hat{\mu}_1/\sigma_1\right\vert,\left\vert\hat{\mu}_2/\sigma_2\right\vert\} \geq \Phi^{-1}(1-\alpha)\}.$$
   \end{proposition}

Whenever $c_\alpha=\Phi^{-1}(1-\alpha)$, the recommended test's rejection region coincides with $\bm{R}_\alpha$. Hence, in this case, the recommended test is uniformly most powerful among monotonic, valid tests. In particular, this conclusion provably holds for $\rho \geq 0$ and numerically obtains for most values of $\rho$, e.g., for $\rho>-.8$ when $\alpha=.05$. We conclude that the test recommended here is in a strong sense optimal for most practically relevant parameter values. For the special case of strong negative correlation of estimators, we do not recommend it because, for example, a moment selection test might fare better \citep[see, e.g.,][]{Kim23}.

\section{More General Null Hypotheses}\label{sec:general}
This section sketches out a generalization of our results that is motivated by analysis in \citetalias{BMW}.
They consider a class of hypotheses where $(\mu_1,\mu_2)$ may not be constrained to the first and third quadrant but to either the positive or the negative span of two positive vectors (see Appendix C of \citetalias{BMW}). If these vectors are known, then this is the generalization of the previous problem to null hypotheses that are the union of a polyhedral cone and its negative (with apex at the origin, though the origin itself is arbitrary).

The insight behind this generalization is that any such cone can be expressed as union of the positive and negative quadrants upon a base change. Simply, if a  full dimensional cone $C$ is spanned by vectors $(b_1,b_2)$, then $\bm{A}=[b_1 ~~ b_2]^{-1}$ denotes the base change matrix under which $C$ is the positive orthant. The null hypothesis $$H_0:(\mu_1,\mu_2) \in C \cup -C$$ is then equivalent to 
$$H_0:(\nu_1,\nu_2) \equiv (\bm{A}\mu_1,\bm{A}\mu_2) \in \mathbb{R}_+^2 \cup  \mathbb{R}_-^2.$$ 

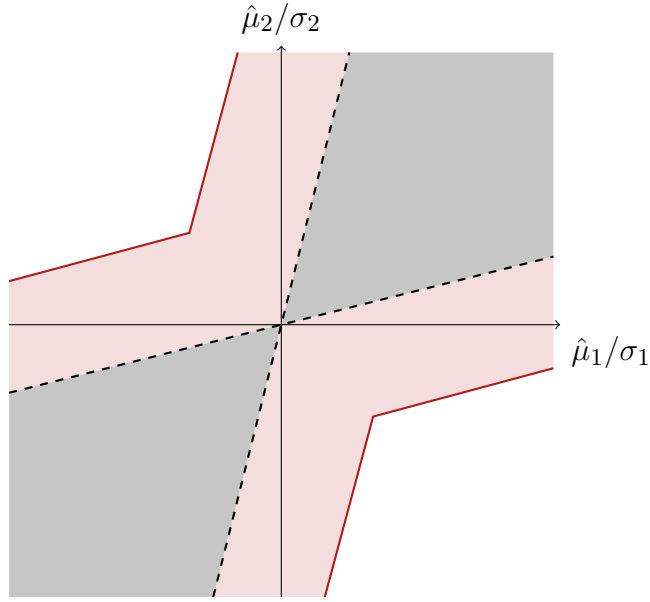
\begin{figure}[t]
\tikzstyle{line} = [draw, -latex']
\begin{tikzpicture}[node distance = 2cm, auto, scale=.9]
\linespread{0.9}
\path [fill=cornellred!15] (0,0) -- (4,1) -- (4,-0.65)-- (1.35,-1.35) -- (0.65,-4) -- (-1,-4);
\draw[cornellred,thick] (4,-0.64)-- (1.35,-1.35) -- (0.64,-4);
\path [fill=cornellred!15] (0,0) -- (1,4) -- (-.65,4)-- (-1.35,1.35) -- (-4,.65) -- (-4,-1);
\draw[cornellred,thick] (-.64,4)-- (-1.35,1.35) -- (-4,.64);
\path [fill=darkgray!30] (0,0) -- (-1,-4) -- (-4,-4)-- (-4,-1);
\path [fill=darkgray!30] (0,0) -- (1,4) -- (4,4)-- (4,1);
\draw[dashed,thick] (-1,-4)-- (1,4) ;
\draw[dashed,thick] (-4,-1)-- (4,1) ;
\draw[->] (-4,0) -- (4.1,0) node[below right] {$\hat \mu_1/\sigma_1$};
\draw[->] (0,-4) -- (0,4.1) node[above] {$\hat \mu_2/\sigma_2$};\end{tikzpicture}
\caption{Visualization of the recommended test for more general null hypotheses; cf. Section \ref{sec:general}.}
\label{fig:general}
\end{figure}
The methods developed here apply, where we have that
\begin{equation*}
    (\hat{\nu}_1,\hat{\nu}_2) \sim \mathcal{N} \left(0,\bm{A}  \begin{pmatrix}\sigma^2_1 & \rho\sigma_1\sigma_2 \\ \rho\sigma_1\sigma_2 & \sigma^2_2\end{pmatrix} \bm{A}'\right).
\end{equation*}
To illustrate this numerically, consider the cone that is spanned by $\{(2,1),(1,2)\}$. Then
\begin{equation*}
    \bm{A}=  \begin{bmatrix} 2~~~~ & 1 \\ 1~~~~ & 2 \end{bmatrix}^{-1} =\frac{1}{3} \begin{bmatrix} 2~~ & -1 \\ -1~~ & 2 \end{bmatrix}  \implies \begin{bmatrix} \nu_1 \\ \nu_2 \end{bmatrix} = \begin{bmatrix} 2\mu_1/3-\mu_2/3 \\ -\mu_1/3 + 2\mu_2/3 \end{bmatrix}.
\end{equation*}
Therefore,
$$(\mu_1,\mu_2) \in C \cup -C   ~~~\Longleftrightarrow ~~~ \nu_1 \cdot \nu_2 \geq 0,$$
and the previous analysis applies. The asymptotic version of this analysis can also easily accommodate the possibility that $(\nu_1,\nu_2)$ are approximated by consistent estimators $(\hat{\nu}_1,\hat{\nu}_2)$, as is the case in \citetalias{BMW}. The resulting test's nonrejection region is visualized in Figure \ref{fig:general}.

\section{Empirical Applications}
A simple application of our results revisits the papers whose analyses motivated our work, \citet{Kowalski} and \citetalias{BMW}, and assesses statistical significance for a mediation analysis exercise carried out in \citet{DippelEtAl2021}.
We report our results in Table \ref{table_results}.

Both \citetalias{BMW} and \citet{Kowalski} report Bonferroni adjusted p-values, and it follows from our results that these p-values can be halved; see the second block of the column labeled Kowalski in Table \ref{table_results}.
\citet[pp.~448-449]{Kowalski} also reports a p-value of $0.023$ using her heuristic bootstrap procedure, whereas our preferred test yields a p-value of $0.0215$; note how these p-values are ordered in accordance with Proposition \ref{prop:compare}. 

\begin{table}
\begin{threeparttable}
\caption{Empirical Examples}
\label{table_results}

\begin{tabular}{lccc}

\toprule

    &  \textbf{Kowalski} & \textbf{Dippel et. al (1)} & \textbf{Dippel et. al. (2)} \\
\toprule

$\hat{\mu}_1$ & -301$^1$ & -0.024 & -0.273   \\
   & [N/A]  & [0.004] & [0.001] \\
$\hat{\mu}_2$ & 148$^1$ & -3.927 & -0.582   \\
   & [N/A]  & [0.030] & [0.025] \\

Reported product & -44,311$^1$ & 0.094 & 0.159 \\

$\rho$ & 0$^2$ & 0.219$^3$ & 0.038$^3$ \\

\hline
P-values \\
\hspace{.25cm}In original paper & 0.043 (BMW) & (none) & (none) \\
    & 0.023 (Heur. Boot.$^4$) & & \\
    \\
\hspace{.25cm}Our method & 0.0215 & 0.015 & 0.013 \\

\hline

Location in original paper & Fig. 5 & Table 5, Col 1 & Table 5, Col 2 \\
 & \& text p. 448-449 & \& Table 4, Col 5 &  \& Table 4, Col 5  \\

Key parameter estimate & Eq. 2 & $\beta^T_M \cdot \beta^M_Y$ & $\beta^T_M \cdot \beta^M_Y$ \\

\# observations & 19,505 & 730 & 730 \\
\# clusters &  & 93 & 93 \\

\hline 
\end{tabular}
    \begin{tablenotes}
      \small
      \item Table notes: $(1)$ \citeauthor{Kowalski}'s Figure 5 reports $\hat{\mu}_2$ as $206-58$; the note to that figure indicates that some differences between statistics might not appear internally consistent because of rounding. (2)  See Proposition \ref{prop:kowalski}. (3) bootstrap-based calculations. (4) ``Heur. Boot.'' stands for heuristic bootstrap as described in Eq.~\eqref{heur_boot}.
    \end{tablenotes}
\end{threeparttable}
\end{table}

\cite{DippelEtAl2021} examine the mediation role that labor market adjustments play in the impact of trade exposure on far-right party support.  A key parameter of interest in this study is the product of $\beta_M^T$ (the effect of trade exposure on local labor markets) and $\beta_Y^M$ (the effect of the strength of the labor market on far-right party support), where each piece comes from a separate estimation model. The sign of this product gives the direction of the mediation effect. It is positive, but \citet{DippelEtAl2021} do not remark on significance. We next verify that the indirect effect is, in fact, statistically significantly positive.

The parameters are estimated on the same data, and their estimating equations have some variables in common, so $\hat{\beta}_M^T$ and $\hat{\beta}_Y^M$ may be correlated. Practical complications in estimating this correlation include the facts that each model is estimated via 2SLS and that the data are grouped by 93 commuting zones, meaning that cluster-robust inference is warranted. 
We accordingly perform a cluster block bootstrap (with 10,000 replications).  We sample over clusters defined by commuting zones, re-estimate the parameters within each bootstrap sample, and compute the resulting correlation between the two estimators.\footnote{We employ a trimmed estimator of bootstrap covariance as outlined in \citet[][Section 10.14]{Hansen}. Specifically, we set to zero the top $1\%$ of bootstrap samples ordered by the bootstrap analog of $\Vert(\hat{\beta}_M^T,\hat{\beta}_Y^M)\Vert$.}

We focus on specifications (1) and (2) reported in Table 5 of \cite{DippelEtAl2021}. The difference between them lies in the mediating variables, with (1) defining the mediator as change in log employment and (2) defining the mediator as an aggregated labor market component derived from principal component analysis. Based on our estimation, the correlation between estimators reported in Column 1 of Table 5 in \citet{DippelEtAl2021} is 0.219; for the estimators in Column 2, it is 0.038. Because we can confidently assess both correlations to be greater than zero (and certainly greater than $-.8$), the one-sided critical value applies; similarly, the p-value of the test in Eq. \eqref{eq:simple_case} can be obtained as if the test were simple one-sided.

Again, \citet[Table 5, Panel B]{DippelEtAl2021} find that $\hat{\beta}_M^T\hat{\beta}_Y^M>0$ and therefore obtain a positive point estimate for the amount of mediation, but they do not test the hypothesis of no or negative mediation and accordingly cannot (and do not) claim that their estimate is statistically significant. We can now clarify that such a claim is indeed warranted.  
For example, given that $\hat{\beta}_M^T\hat{\beta}_Y^M>0$, at significance level $\alpha=0.05$ we can compare $\min(|\hat{t}_M^T|,|\hat{t}_Y^M|)$ to $1.645$, where $\hat{t}_M^T=\hat{\beta}_M^T/\hat{\sigma}_M^T$ and $\hat{t}_Y^M=\hat{\beta}_Y^M/\hat{\sigma}_Y^M$. Using \citeauthor{DippelEtAl2021}'s (\citeyear{DippelEtAl2021}) replication package to compute these numbers, we conclude that the mediator plays a statistically significant positive role in the overall impact. Indeed, with the (bootstrapped) knowledge that the correlation between the two parameters is non-negative, the overall p-value against zero mediation can be immediately read off their Tables 4 and 5, which report p-values from (two-sided) tests of significance for $(\hat{\beta}_M^T,\hat{\beta}_Y^T)$. The recommended test's p-value is just one half of the larger of these, i.e. $0.015$ for specification (1) and $0.013$ for specification (2).

\section{Conclusion}
We provide simple tests for bivariate sign congruence, a problem that has recently received renewed attention in empirical research and has a long history in the literature on testing for heterogeneous treatment effects. The problem is hard because the null hypothesis is irregularly shaped at the origin of parameter space. We rediscover a test from an older statistics literature that unambiguously improves on recent empirical work in the social sciences, and we clarify when some of these other proposals apply. We argue that, for most parameter values, the test is optimal subject to some arguably plausible desiderata; also, we illustrate why some such desiderata are necessary. The results can be immediately applied to existing empirical work, e.g. by dividing p-values reported in some papers by $2$. We emphasize that our results' striking simplicity owes to restricting attention to the bivariate testing problem. For $K>2$ parameters with uncorrelated estimators, we recommend the likelihood ratio test in \citet{GS85}; for the most general case with unrestricted covariance matrix, we refer to \citet{Kim23}.

\begin{appendix}
\section{Proofs}\label{appendix}
The following lemma will be used repeatedly.
\begin{lemma}\label{lem:andersen}
    Let \eqref{eq:normal_est} hold.
     Consider any test whose rejection region can be expressed as interior of a set $RR$ defined as 
$$RR \equiv \bigcup_{z \in \mathbb{R}} RR(z),~~~~ RR(z) \equiv  \left\{\begin{bmatrix}-z \\ z \end{bmatrix} + a \begin{bmatrix}1 \\ 1\end{bmatrix}:a \in [-a^*(z),a^*(z)]\right\}$$
    for some function $a^*(\cdot):\mathbb{R} \to [0,\infty]$. In words, the intersection of $RR$ with any ascending $45^\circ$-line is a (possibly unbounded) line segment centered at the descending $45^\circ$-line.

Then on $H_0$ as defined in Eq.~\eqref{eq:simple_case}, the supremum of the test's rejection probability (i.e., its size) equals its supremum on the null's boundary.
\end{lemma}

\begin{proof}
The null hypothesis in Eq.~\eqref{eq:simple_case} can equivalently be written as
\begin{align*}
    H_0':\frac{\mu_1}{\sigma_1}\cdot\frac{\mu_2}{\sigma_2}\ge 0.
\end{align*}
For the remainder of this proof, whenever talking about the null hypothesis we refer to $H_0'$, and to simplify expressions we set $\sigma_1=\sigma_2=1$.
For any fixed $(\mu_1,\mu_2)$, we define
\begin{eqnarray*}
    a = \frac{\mu_1+\mu_2}{2}, ~~~  \lambda = \frac{\mu_2-\mu_1}{2},
\end{eqnarray*}
and we reparametrize  $(\mu_1,\mu_2)=(-\lambda+a,\lambda+a)$. We show below that, for any fixed $\lambda$, rejection probability is nonincreasing in $|a|$. Since no point in the interior of $H_0'$ corresponds to $a=0$, it follows that for any interior point, there is a point on the boundary of $H_0'$ that induces at least the same rejection probability. This establishes the lemma.

To see the point about rejection probability, write
\begin{equation*}
    \begin{pmatrix}\hat{\mu}_1 \\ \hat{\mu}_2 \end{pmatrix} \sim \mathcal{N}\left(\begin{pmatrix}-\lambda+a \\ \lambda+a \end{pmatrix},\begin{pmatrix}1 & ~~~\rho \\ \rho & ~~~1\end{pmatrix}\right)
\end{equation*}
and note for future reference that
\begin{equation}
    \begin{pmatrix}\hat{\mu}_1+\hat{\mu}_2 \\ \hat{\mu}_1-\hat{\mu}_2 \end{pmatrix} \sim \mathcal{N}\left(\begin{pmatrix}2a \\ -2\lambda \end{pmatrix},\begin{pmatrix}2(1+\rho) & 0 \\ 0 & 2(1-\rho)\end{pmatrix}\right). \label{eq:coord_change}
\end{equation}
The claim is that, for any fixed $\lambda$, $\Pr((\hat{\mu}_1,\hat{\mu}_2)\in RR)$ is nonincreasing in $|a|$. Indeed, we can write
\begin{eqnarray*}
    && \Pr((\hat{\mu}_1,\hat{\mu}_2) \in RR) \\
    &=& \int \Pr((\hat{\mu}_1,\hat{\mu}_2) \in RR(z)\mid \hat{\mu}_1-\hat{\mu}_2=2z) \phi\left(\frac{2z+2\lambda}{\sqrt{2(1-\rho)}}\right)dz \\
     &=& \int \Pr(\hat{\mu}_1+\hat{\mu}_2 \in [-a^*(z),a^*(z)]\mid \hat{\mu}_1-\hat{\mu}_2=2z) \phi\left(\frac{2z+2\lambda}{\sqrt{2(1-\rho)}}\right)dz \\
      &=& \int \left[\Phi\left(\frac{-2a+a^*(z)}{\sqrt{2(1+\rho)}}\right)-\Phi\left(\frac{-2a-a^*(z)}{\sqrt{2(1+\rho)}}\right)\right] \phi\left(\frac{2z+2\lambda}{\sqrt{2(1-\rho)}}\right)dz.
\end{eqnarray*}
Here, all steps use probability calculus, the definition of $RR$, and \eqref{eq:coord_change}. Note in particular that the conditioning event in the first step is such that $RR(z)$ is the only relevant slice of $RR$. Taking derivatives in the last line (or applying Anderson's Lemma in the second last line, again also using \eqref{eq:coord_change}), one can see that $\Pr(\hat{\mu}_1+\hat{\mu}_2 \in [-a^*(z),a^*(z)]| \hat{\mu}_1-\hat{\mu}_2=2z)$ is nonincreasing in $|a|$ for every $z$, establishing the claim. Finally, the boundary of $RR$ has probability $0$, so the conclusion extends to its interior.
\end{proof}

\subsection{Proof of Theorem \ref{th:main}} \label{app:proof}
Without loss of generality (see Lemma \ref{lem:andersen}), set $\sigma_1=\sigma_2=1$. Lemma \ref{lem:andersen} applies to the recommended test, hence its rejection probability is maximized on the boundary of the null. Considering also the setup's and the test's symmetry, it suffices to consider the part of the boundary defined by $\mu_1=0$. But then the test's size can be written as (using $\vee$ for ``or'' and $\wedge$ for ``and'') 
\begin{eqnarray}
    \sup_{\mu_1=0,\mu_2 \geq 0} \Pr(\hat{\mu}_1 \hat{\mu}_2<0 \wedge |\hat{\mu}_1| > c_\alpha \wedge |\hat{\mu}_2|> c_\alpha), \label{eq:supremum}
\end{eqnarray}
which equals $\alpha$ by \eqref{eq:def:c_alpha}-\eqref{eq:def:c_alpha2}, identifying $(X_1,X_2)$ there with $(\hat{\mu}_1-\mu_1,\hat{\mu}_2-\mu_2)$ here. This establishes all claims except validity of $\Phi^{-1}(1-\alpha)$ for $\rho \geq 0$. To see this last claim, suppose $\rho \geq 0$ and write 
\begin{eqnarray*}
&& \Pr(\hat{\mu}_1 \hat{\mu}_2<0 \wedge |\hat{\mu}_1| > c_\alpha \wedge |\hat{\mu}_2|> c_\alpha) \\
&=& \Pr\bigl([\hat{\mu}_1<-\Phi(1-\alpha) \wedge \hat{\mu}_2 > \Phi(1-\alpha)] \vee [\hat{\mu}_1>\Phi(1-\alpha) \wedge \hat{\mu}_2 <- \Phi(1-\alpha)]\bigr) \\
&=& \Pr\bigl([X_1<-\Phi(1-\alpha) \wedge \mu_2+X_2 > \Phi(1-\alpha)] \vee [X_1>\Phi(1-\alpha) \wedge \mu_2+X_2 <-\Phi(1-\alpha)]\bigr) \\
&\leq &  \Pr\bigl([X_1<-\Phi(1-\alpha) \wedge \mu_2+X_2 > \Phi(1-\alpha)] \vee [X_1<-\Phi(1-\alpha) \wedge \mu_2+X_2 <-\Phi(1-\alpha)]\bigr) \\
&<& \Pr \bigl(X_1<-\Phi(1-\alpha)\bigr) = \alpha.
\end{eqnarray*}
Here, the second equality again identifies $(\hat{\mu}_1-\mu_1,\hat{\mu}_2-\mu_2)$ with $(X_1,X_2)$ from \eqref{eq:def:c_alpha2}; the weak inequality uses that $\rho \geq 0$ and the distribution of $(X_1,X_2)$ is therefore characterized by positive quadrant dependence; and the final inequality is elementary probability calculus. Despite the strict inequality, the test is not conservative because the supremum in \eqref{eq:supremum} can be approximated by letting $\mu_2 \to \infty$. 

\subsection{Proof of Proposition \ref{th:feasible}}
Recall the claim is that
\begin{eqnarray*}
   \lim_{n \to \infty}\sup_{\mu_1 \mu_2 \geq 0}\Pr(\hat{\mu}_1 \hat{\mu}_2<0 \wedge \sqrt{n}|\hat{\mu}_1|/\hat{\sigma}_1\geq \hat{c}_\alpha \wedge \sqrt{n}|\hat{\mu}_2|/\hat{\sigma}_2\geq \hat{c}_\alpha)=\alpha. 
\end{eqnarray*}
where it is understood that the supremum is over data generating processes that fulfill Assumption \ref{as:asy} (as well as $\mu_1\mu_2\geq 0$). By definition of a supremum, there exists a sequence of such processes characterized by parameters $(\mu_{n1},\mu_{n2},\sigma_{n1},\sigma_{n2},\rho_n)_{n=1}^\infty$ s.t. (for clarity, estimators inherit $n$-subscripts) 
\begin{multline*}
   \lim_{n \to \infty}\Bigl\{\sup_{\mu_1 \mu_2 \geq 0}\Pr(\hat{\mu}_1 \hat{\mu}_2<0 \wedge \sqrt{n}|\hat{\mu}_1|/\hat{\sigma}_1\geq \hat{c}_\alpha \wedge \sqrt{n}|\hat{\mu}_2|/\hat{\sigma}_2\geq \hat{c}_\alpha)\Bigr. \\
    -  \Bigl. \Pr(\hat{\mu}_{n1} \hat{\mu}_{n2}<0 \wedge \sqrt{n}|\hat{\mu}_{n1}|/\hat{\sigma}_{n1}\geq \hat{c}_\alpha \wedge \sqrt{n}|\hat{\mu}_{n2}|/\hat{\sigma}_{n2}\geq \hat{c}_\alpha)\Bigr\}=0.
\end{multline*}
It therefore suffices to show that, along any arbitrary such sequence, 
\begin{eqnarray}
 \lim_{n \to \infty}\Pr(\hat{\mu}_{n1} \hat{\mu}_{n2}<0 \wedge \sqrt{n}|\hat{\mu}_{n1}|/\hat{\sigma}_{n1}\geq \hat{c}_\alpha \wedge \sqrt{n}|\hat{\mu}_{n2}|/\hat{\sigma}_{n2}\geq \hat{c}_\alpha) \leq \alpha \label{eq:proof_target}
\end{eqnarray}
and that the inequality binds for some such sequence. Importantly, from the uniformity in Assumption \ref{as:asy}, we can claim that $\hat{\mu}_{n1}-\mu_n\overset{p}{\to}0$ etc.

Suppose first that $\sqrt{n}\mu_{n1}/\sigma_{n1}\to \infty$. Then \eqref{eq:proof_target} simplifies to
\begin{eqnarray*}
 \lim_{n \to \infty}\Pr(\sqrt{n}\hat{\mu}_{n2}/\hat{\sigma}_{n2}\leq -\hat{c}_\alpha).
\end{eqnarray*}
As $\mu_{n1}>0$ implies $\mu_{n2}\geq 0$, this limit equals at most $\alpha$, with equality if $\mu_{n2}=0$. The analogous argument applies to $\sqrt{n}\mu_{n1}/\sigma_{n1}\to -\infty$ and $\sqrt{n}\mu_{n2}/\sigma_{n2}\to \pm\infty$. In all other cases, $(\sqrt{n}\mu_{n1}/\sigma_{n1},\sqrt{n}\mu_{n2}/\sigma_{n2},\rho_n)_{n=1}^\infty$ has finitely many accumulation points; by extracting subsequences if necessary, it suffices to consider sequences s.t.
\begin{eqnarray*}
    (\sqrt{n}\mu_{n1}/\sigma_{n1},\sqrt{n}\mu_{n2}/\sigma_{n2})\to (\gamma_1,\gamma_2)\in \{\mathbb{R}^2:\gamma_1 \gamma_2 \geq 0\} \text{  and  } \rho_n \to \rho \in [-1,1].
\end{eqnarray*}
Fix any such sequence, define $X_{nj}=\sqrt{n}(\hat{\mu}_{nj}-\mu_{nj})/\sigma_{nj}$, and write
\begin{eqnarray*}
&&   \lim_{n \to \infty}\Pr(\hat{\mu}_{n1} \hat{\mu}_{n2}<0 \wedge \sqrt{n}|\hat{\mu}_{n1}|/\hat{\sigma}_{n1}\geq \hat{c}_\alpha \wedge \sqrt{n}|\hat{\mu}_{n2}|/\hat{\sigma}_{n2}\geq \hat{c}_\alpha) \\
&=& \lim_{n \to \infty}\Pr\bigl(\tfrac{1}{n}(\sqrt{n}\mu_{n1}+\sigma_{n1}X_{n1})(\sqrt{n}\mu_{n2}+\sigma_{n2}X_{n2})<0 \bigr. \\
&& ~~~~\bigl.\wedge \left\vert\sqrt{n}\mu_{n1}+ \sigma_{n1}X_{n1}\right\vert /\hat{\sigma}_{n1} \geq \hat{c}_\alpha \wedge \left\vert\sqrt{n}\mu_2+\sigma_{n2}X_{n2}\right\vert /\hat{\sigma}_{n2} \geq \hat{c}_\alpha\bigr) \\
&=& \lim_{n \to \infty}\Pr\bigl((\sqrt{n}\mu_{n1}/\sigma_{n1}+X_{n1})(\sqrt{n}\mu_{n2}/\sigma_{n2}+X_{n2})<0 \bigr. \\
&& ~~~~\bigl.\wedge \left\vert\sqrt{n}\mu_{n1}/\sigma_{n1} + X_{n1}\right\vert \cdot \sigma_{n1}/\hat{\sigma}_{n1} \geq \hat{c}_\alpha \wedge \left\vert\sqrt{n}\mu_2/\sigma_{n2}+X_{n2}\right\vert \cdot\sigma_{n2}/\hat{\sigma}_{n2} \geq \hat{c}_\alpha\bigr) \\
&=& \Pr\left((\gamma_1+X_1)(\gamma_2+X_2)<0 \wedge \left\vert\gamma_1+ X_1\right\vert  \geq c_\alpha \wedge \left\vert\gamma_2+X_2\right\vert  \geq c_\alpha\right) \\
&\leq & \sup_{\gamma_1,\gamma_2:\gamma_1 \gamma_2 \geq 0} \Pr\left((\gamma_1+X_1)(\gamma_2+X_2)<0 \wedge \left\vert\gamma_1+ X_1\right\vert  \geq c_\alpha \wedge \left\vert\gamma_2+X_2\right\vert  \geq c_\alpha\right) \\
&=& \alpha.
\end{eqnarray*}
Here, the first step substitutes in for $X_{nj}$. The second step rearranges terms and uses that inequalities can be divided through by positive constants. The next step takes limits; note in particular that $\hat{\rho}_n \overset{p}{\to}\rho$ and therefore both $(X_{n1},X_{n2})\overset{d}{\to}(X_1,X_2)$, where $(X_1,X_2)$ is as in \eqref{eq:def:c_alpha2}, and $\hat{c}_\alpha\overset{p}{\to}c_\alpha$, where $c_\alpha$ is as in \eqref{eq:def:c_alpha}. The inequality step uses that $\gamma_1 \gamma_2 \geq 0$. The last step invokes Theorem \ref{th:main}, identifying $(\mu_j,\hat{\mu}_j)$ there with $(\gamma_j,\gamma_j+X_j)$ here. The additional claim when $\rho \geq 0$ is established just as before.

\subsection{Proof of Proposition \ref{prop:heuristic}}
One can easily verify that the rejection probability in question approaches $1$ as $(\mu_1,\mu_2,\sigma_1,\sigma_2,\rho) \to (0,0,1,1,-1)$, whereas it approaches $0$ as $(\mu_1,\mu_2,\sigma_1,\sigma_2,\rho) \to (0,0,1,1,1)$. This is irrespective of whether $(\mu_1,\mu_2)$ approach their limits from inside or outside of $H_0$.

\subsection{Proof of Proposition \ref{prop:compare}}
We again set $\sigma_1=\sigma_2=1$. For the \citetalias{BMW} test, the claim is obvious (for any $\rho$). For the heuristic bootstrap, fix $\rho=0$ and consider any $(\hat{\mu}_1,\hat{\mu}_2)$ on the boundary of our test's rejection region. Without further loss of generality, set $\hat{\mu}_1=\Phi(1-\alpha)>0$. Defining $(X_1^*,X_2^*)=(\hat{\mu}_1^*-\hat{\mu}_1,\hat{\mu}_2^*-\hat{\mu}_2)$ and observing that $$ \left(\begin{pmatrix}X_1^* \\ X_2^*\end{pmatrix}\bigm\vert \hat{\mu}_1,\hat{\mu}_2\right) \sim \mathcal{N}\left(0,\begin{pmatrix}1 & ~~~ 0 \\ 0 & ~~~ 1 \end{pmatrix}\right),$$ one can write (the conditioning event is unchanged throughout the display)
\begin{eqnarray*}
\Pr(\hat{\mu}_1^* \cdot \hat{\mu}_2^* >0\mid \hat{\mu}_1,\hat{\mu}_2) &=& \Pr((\hat{\mu}_1+X_1^*)(\hat{\mu}_2+X_2^*) >0\mid\cdot~) \\
&=& \Pr\bigl([X_1^* > -\hat{\mu}_1 \wedge X_2^* > -\hat{\mu}_2]\vee [X_1^* < -\hat{\mu}_1 \wedge X_2^* < -\hat{\mu}_2]\mid\cdot~\bigr) \\
&>& \Pr\bigl([X_1^* < -\hat{\mu}_1 \wedge X_2^* > -\hat{\mu}_2]\vee [X_1^* < -\hat{\mu}_1 \wedge X_2^* < -\hat{\mu}_2]\mid\cdot~\bigr) \\
&=& \Pr\bigl(X_1^* < -\hat{\mu}_1\mid\cdot~\bigr)=1-\alpha.
\end{eqnarray*}
It follows that the entire boundary of the recommended test's rejection region is in the interior of the heuristic bootstrap test's \textit{nonrejection} region.

To see that the \citetalias{BMW} rejection region is interior to the heuristic bootstrap one, consider maximizing the heuristic bootstrap p-value of $(\hat{\mu}_1,\hat{\mu}_2)$ on the \citetalias{BMW} rejection region. It is easy to see that this problem is solved by $(\Phi(1-\alpha/2),-\Phi(1-\alpha/2))$. But at that point, the heuristic bootstrap p-value equals $2 \cdot \alpha/2 - (\alpha/2)^2 < \alpha$, so that it is interior to the heuristic bootstrap rejection region.

\subsection{Proof of Proposition \ref{prop:kowalski}}
In this proof only, all notation is exactly as in \citet{Kowalski}. Substitute
\begin{eqnarray*}
E(Y_T\mid A) &=& E(Y\mid D=1,Z=0) \\
E(Y_U\mid N) &=& E(Y\mid D=0,Z=1) \\
E(Y_T\mid C) &=& \tfrac{p_I}{p_I-p_C}E(Y\mid D=1,Z=1)-\tfrac{p_C}{p_I-p_C}E(Y\mid D=1,Z=0) \\
E(Y_U\mid C) &=& \tfrac{1-p_C}{p_I-p_C}E(Y\mid D=0,Z=0)-\tfrac{1-p_I}{p_I-p_C}E(Y\mid D=0,Z=1) 
\end{eqnarray*}
into 
  $$ \tau \equiv \bigl(E(Y_U\mid C)-E(Y_U\mid N)\bigr)\cdot\bigl(E(Y_T\mid A)-E(Y_T\mid C)\bigr) $$
to find
\begin{eqnarray*}
\tau &=& \bigl(\tfrac{1-p_C}{p_I-p_C}E(Y\mid D=0,Z=0)-\tfrac{1-p_I}{p_I-p_C}E(Y\mid D=0,Z=1)-E(Y\mid D=0,Z=1) \bigr) \\
&& \cdot~\bigl(E(Y\mid D=1,Z=0)-\tfrac{p_I}{p_I-p_C}E(Y\mid D=1,Z=1)+\tfrac{p_C}{p_I-p_C}E(Y\mid D=1,Z=0)\bigr) \\
&=& \tfrac{1-p_C}{p_I-p_C}\bigl(E(Y\mid D=0,Z=0)-E(Y\mid D=0,Z=1) \bigr) \\
&& \cdot~\tfrac{p_I}{p_I-p_C}\bigl(E(Y\mid D=1,Z=0)-E(Y\mid D=1,Z=1)\bigr) \\
&= & \tfrac{p_I(1-p_C)}{(p_I-p_C)^2}\bigl(E(Y\mid D=0,Z=0)-E(Y\mid D=0,Z=1) \bigr)\\
&& \cdot~\bigl(E(Y\mid D=1,Z=0)-E(Y\mid D=1,Z=1)\bigr),
\end{eqnarray*}
so that it suffices to sign
\begin{multline*}
   \bigl((E(Y\mid D=0,Z=0)-E(Y\mid D=0,Z=1)\bigr)\cdot\bigl((E(Y\mid D=1,Z=0)-E(Y\mid D=1,Z=1)\bigr), 
\end{multline*}
but these expectations condition on mutually exclusive events and their sample analogs are computed from mutually exclusive subsamples.

\subsection{Proof of Proposition \ref{prop:fractal}}
Again set $\sigma_1=\sigma_2=1$. To see uniform validity, consider first any parameter value $(\mu,0)$. 
Let $\phi(\cdot)$ be the indicator function for the test in Eq.~\eqref{eq:fractal_test}.
Then for any $m_1 \neq 0$, it is easily seen that $\Pr(\phi(\hat{\mu}_1,\hat{\mu}_2)=1\mid\hat{\mu}_1=m_1)=.05$. In particular, consider that $\hat{\mu}_2$ is standard normal and that conditionally on any $\hat{\mu}_1\neq 0$, the probability of $(\hat{\mu}_1,\hat{\mu}_2)$ falling in the rejection region is the probability of $\hat{\mu}_2$ hitting some $.05$-quantile bin of the standard normal distribution. The argument for any parameter value $(0,\mu)$ is analogous.

To see unbiasedness, observe that Lemma \ref{lem:andersen} applies; hence, for any fixed value of $(\mu_2-\mu_1)$, probability of rejection strictly decreases in $\left\vert \mu_1+\mu_2 \right\vert$. This establishes the remainder of the proposition's claims.

\subsection{Proof of Proposition \ref{prop:monotonic}}
Suppose by contradiction that the rejection region contains a point $(m_1,m_2)$ with $m_1>0$ and $m_2>-\Phi^{-1}(1-\alpha)$. Then by monotonicity, it also contains $(m_1+\lambda,m_2)$ for any $\lambda \geq 0$. But this implies that, along any sequence of parameter values $\{(0,\mu_{1,n}):\mu_{1,n} \to \infty\}$, rejection probability converges to (at least) $\Phi(-m_2)>\alpha$. The argument for other possible values of $(m_1,m_2)$ is similar.

\section{Highly Accurate Critical Values}\label{app:critical_values}

\begin{table}
\begin{threeparttable}
\caption{Critical values computed to high accuracy}
\label{table:critical_values}

\begin{tabularx}{\textwidth}{XXXX}
\toprule
$\bm{\rho}$ & $\bm{\alpha = 0.1}$ & $\bm{\alpha = 0.05}$ & $\bm{\alpha = 0.01}$ \\ 
\midrule
$-1$ & 1.64485363 & 1.95996398 & 2.57582930 \\
$-0.95$ & 1.50530474 & 1.81773816 & 2.42829856 \\
$-0.90$ & 1.43932899 & 1.74893328 & 2.35384002 \\
$-0.85$ & 1.38524452 & 1.69190984 & 2.32671651 \\
$-0.80$ & 1.33720245 & 1.64878158 & 2.32634836 \\
$-0.75$ & 1.29315300 & 1.64488267 & 2.32634787 \\
$-0.70$ & 1.28170171 & 1.64485364 & 2.32634787\textsuperscript{*} \\
$-0.65$ & 1.28155170 & 1.64485363 & 2.32634787\textsuperscript{*} \\
$-0.60$ & 1.28155157 & 1.64485363\textsuperscript{*} & 2.32634787\textsuperscript{*} \\
$-0.55$ & 1.28155157\textsuperscript{*} & 1.64485363\textsuperscript{*} & 2.32634787\textsuperscript{*} \\
$0$     & 1.28155157 & 1.64485363 & 2.32634787 \\
\bottomrule

\end{tabularx}

    \begin{tablenotes}
      \small
      \item Table notes: Values with asterisks coincide with one-sided critical values to 15 decimals.
    \end{tablenotes}
\end{threeparttable}
\end{table}

Table \ref{table:critical_values} displays the critical value $c_\alpha$ from \eqref{eq:def:c_alpha} for different values of $\rho$ and $\alpha$. Recall that $c_\alpha$ exactly coincides with the one-sided critical value $\Phi^{-1}(1-\alpha)$for $\rho \geq 0$; the entries for $\rho=0$ in the table replicate these values. It is also easy to show that $\rho=-1$ necessitates the use of two-sided critical values, which are displayed in the first row. Values that are marked with asterisks coincide with one-sided critical values to $15$ decimals. This is also true for any value that would appear below such an entry in the table; the table stops at $\rho=-.55$ because all other rows would read the same. Rejection probabilities were computed in MATLAB using numerical integration via the native \texttt{mvncdf} command; they were evaluated on a $.001$ grid from $(0,0)$ to $(0,20)$ (and $(0,30)$ in some cases); $c_\alpha$ was determined using a bisection algorithm with $60$ steps and starting at $\Phi^{-1}(1-\alpha)$.

\end{appendix}

\bibliographystyle{ecta-fullname} 
\bibliography{sign_testing}

\end{document}